# PoET: the Paranal solar ESPRESSO Telescope

Nuno C. Santos[1,2]
Alexandre Cabral[3,4]
Inês Leite[3,4]
Alain Smette[5]
Manuel Abreu[3,4]
David Alves[3,4]
Jorge H. C. Martins[1]
Manuel Monteiro[1]
André Silva[1,2]
Bachar Wehbe[3,4]
Jimmy Arancibia[5]
Gerardo Ávila[3]
Stephane Brillant[5]
César Cárdenas[5]
Ricardo Clara[3]
Ricardo Gafeira[6,7]
Daniel Gaytan[5]
Christophe Lovis[8]
Nicolas Miranda[5]
Pedro Moreno[1]
António Oliveira[3,4]
Angel Otarola[5]
Francesco Pepe[8]
Pascual Rojas[5]
Ricardo Schmutzer[5]
Danuta Sosnowska[8]
Pierre van der Heyden[5]
Khaled Al Moulla[1]
Vardan Adibekyan[1]
Alba Barka[1,2]
Susana C. C. Barros[1,2]
Pedro Branco[1,2]
Eduardo Cristo[1]
Yuri Damasceno[1,2]
Olivier Demangeon[1]
William Dethier[1]
João P. Faria[8]
João Gomes da Silva[1]
Eduardo Gonçalves[1,2]
Jennifer P. Lucero[1,2]
José Rodrigues[1,2]
Carmen San Nicolas Martinez[1,2]
Ângela Santos[1,2]
Sérgio Sousa[1]
Pedro T. P. Viana[1,2]

[1] Institute of Astrophysics and Space Sciences, CAUP, Porto, Portugal
[2] Department of Physics and Astronomy, Faculty of Sciences, University of Porto, Portugal
[3] Institute of Astrophysics and Space Science, Faculty of Sciences, University of Lisbon, Portugal
[4] Department of Physics, Faculty of Sciences, University of Lisbon, Portugal
[5] ESO
[6] Institute of Astrophysics and Space Sciences, University of Coimbra, OGAUC Portugal
[7] Department of Physics, University of Coimbra, Portugal
[8] Astronomy Department, University of Geneva, Switzerland

The detection and characterisation of other 'Earths', orbiting other suns, is a bold objective of present-day astrophysics. However, this quest is severely challenged by astrophysical 'noise' from the host stars, whose signatures distort the observed spectra. Motivated by this problem, we are building a dedicated facility, the Paranal solar ESPRESSO Telescope (PoET). PoET will collect solar light and channel it into the ESPRESSO spectrograph, allowing us to use the Sun as a proxy to unambiguously identify and understand the sources of relevant variability in solar-type stars.

## The quest for other Earths

More than 5000 extrasolar planets have been confirmed to date. Present-day discoveries have already allowed one major conclusion: rocky planets seem to be ubiquitous around solar-type stars. Despite the impressive results, no Earth-analogue orbiting a Sun-like star has yet been unambiguously discovered and characterised, even if rocky planets around lower-mass/smaller M-dwarf stars are within reach (for example, Faria et al., 2022; Demangeon et al., 2021).

One of the main battle-horses for the detection and characterisation of exoplanets is high-resolution spectroscopy (for example, Mayor, Lovis & Santos, 2014). Doppler spectroscopy measurements, complemented with transit photometry of planets transiting bright nearby stars (Lissauer et al., 2014), are particularly relevant and allow both their mass and radius to be derived, and thus their mean density. Complementary observations and modelling allow their interior and atmospheres to be probed (for example, Ehrenreich et al., 2020).

Ground-based high-resolution spectrographs such as the Echelle SPectrograph for Rocky Exoplanet and Stable Spectroscopic Observations (ESPRESSO; Pepe et al., 2021), capable of achieving radial velocity (RV) precisions down to 10 cm s$^{-1}$ — the typical amplitude of the signal induced by an Earth-like planet orbiting a Sun-like star — represent relevant steps in this effort. ESPRESSO will be complemented by high-resolution optical and near-infrared spectrographs on ESO's Extremely Large Telescope (ELT), for example with the ArmazoNes high Dispersion Echelle Spectrograph (ANDES; Marconi et al., 2024), whose design is optimised for the detection of exoplanet atmospheres. These instruments will be key to following up Earth-like planets detected by missions such as ESA's PLAnetary Transits and Oscillations (PLATO; Rauer et al., 2024).

## The stellar challenge

In this quest, the greatest challenge to overcome is related to stellar physics, or rather to the astrophysical 'noise' coming from the host stars that distorts the observed spectra.

The physical processes underlying magnetically active features (spots, faculae; for example, Shapiro et al., 2016) produce variations in the observed line profiles and positions. As activity-related features appear and disappear from the stellar disc, they induce spectral variability on timescales typical of the rotational period of the star, as well as of its long-term magnetic cycle. Amplitudes can be as high as several tens or even hundreds of metres per second in RV (for example, Meunier et al., 2017; Faria et al., 2020). Stellar granulation (Dravins, 1982) and p-mode oscillations (for example, Chaplin & Miglio, 2013) are also relevant sources of variability, inducing signals of up to several metres per second, depending on the spectral type (Dumusque et al., 2011). For solar-type stars, granulation signals have timescales from a few hours to days, while for p-modes timescales are of the order of a few minutes. An example of the impact of these processes in the RV of the Sun is shown in Figure 1.

Although different approaches are presently used to model the signals produced by these different phenomena, none has proven to correct RV time-series down to the required precision level (for example,

Haywood et al., 2022). Furthermore, our incomplete knowledge of stellar physics can also severely impact our ability to detect and characterise exoplanet atmospheres, or even produce systematics that are several times stronger than the ones produced by the planetary atmosphere itself (for example, Casasayas-Barris et al., 2021; Dethier & Tessore, 2024; and see Figure 1). The effects of these phenomena on the measurements of transmission spectroscopy can be particularly problematic, and raise questions such as: if Earth were an exoplanet, would we be able to tell whether it was habitable by observing its atmosphere?

### The solar promise

The exoplanet community has recognised that progress in this field requires identifying in detail the different physical processes that drive stellar variations. In this context, the Sun is seen as the ideal target and proxy: it is the only star we can resolve. Dedicated instruments have been built, attached to high-precision spectrographs, to observe the 'Sun-as-a-star' (for example, Zhao et al., 2023). Good examples are the High Accuracy Radial velocity Planet Searcher North (HARPS-N) solar telescope (Dumusque et al., 2021) and its counterpart at the HARPS spectrograph (HELIOS). Overall, these experiments have shed new light on the problem, but also show that our current understanding is still inadequate (for example, Milbourne et al., 2021).

The major drawback of these approaches is the fact that the Sun is observed as a star: only disc integrated spectra are obtained. This precludes a detailed analysis of the individual stellar features responsible for the observed spectral deformations.

### The Paranal solar ESPRESSO Telescope

To find new answers we need to obtain disc-resolved, high-precision spectra of the Sun. Similarly to the best instrumentation used in exoplanet research facilities, an adequate instrument has to offer a) spatially resolved spectroscopy with very high wavelength stability, b) very high spectral resolution ($R = \lambda/\Delta\lambda \sim 200\,000$), to adequately resolve photospheric line asymmetries, and c) extended wavelength coverage, for the simultaneous observation of thousands of spectral lines probing different physical conditions. This can be achieved if we link the ESPRESSO spectrograph to a solar telescope: the Paranal solar Espresso Telescope (PoET[1]; see Figure 2).

In a nutshell (see also Leite et al., 2024), PoET will consist of a 600-mm-diameter telescope designed to point to any resolved region in the solar disc and inject the light into ESPRESSO using optical fibres. It will observe the Sun on different spatial scales: from 55 arcsec in angular diameter, the typical size of a medium-sized sunspot, down to 1 arcsec, the typical spatial scale of one solar granule. Simultaneous full disc integrated observations ('the-Sun-as-a-star') will also be possible using a piggyback pointing telescope.

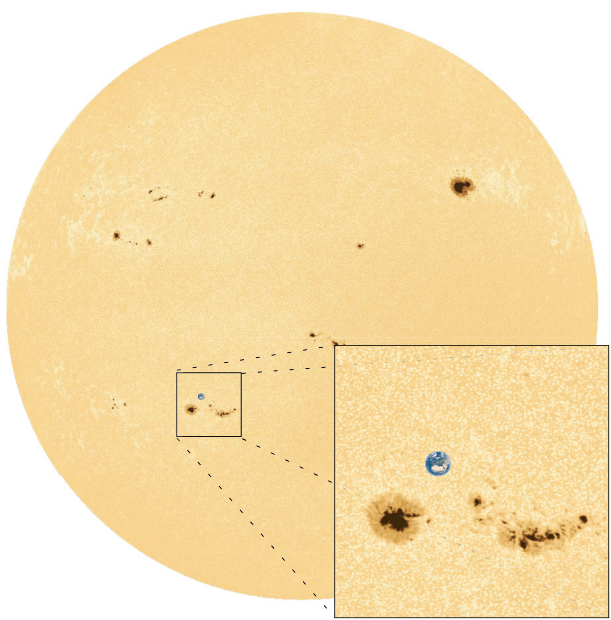
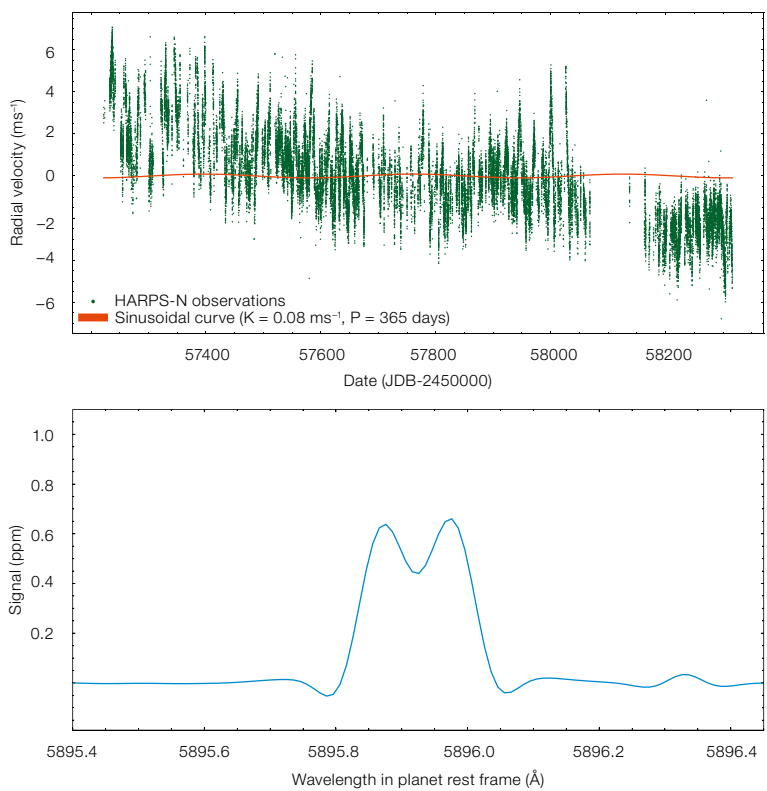

Figure 2. Left: Image of the Sun with Earth superimposed on the same scale. Upper-right: Radial-velocity time-series of the Sun (green points), obtained with the HARPS-N solar telescope (Dumusque et al., 2021), when compared with the expected signature of an Earth-like planet orbiting a Sun-like star (red curve). Lower-right: simulated spurious signature of an absorption spectrum of an atmosphere-less Earth-like planet centered in the sodium line (blue curve), as caused by unaccounted center-to-limb variations. The observed signal has a magnitude level similar to the expected absorption signature of the atmosphere of an Earth-like planet. These plots illustrate the challenges for the detection and characterisation of Earth-like planets orbiting other suns. Image sources: NASA SDO and ESO/M. Kornmesser.

Instrument concept

To achieve simultaneous disc-integrated and (arcsecond-level) disc-resolved observations of the Sun, PoET will consist of a three-telescope system, as shown by the simple schematics in Figure 3.

The main telescope (MT) is from Officina Stellare and its objective is the observation of small areas of the solar disc. It has a Gregorian configuration, chosen because of its intermediate focus. This format, a standard for solar observations, enables the introduction of a heat rejector in the intermediate focal plane, rejecting all the light that falls outside of a 4-arcmin-diameter field. This allows heat to be reduced to less than 2% at the level of the frontend focal plane.

Although seeing-limited, one of PoET's scientific objectives is to perform observations with a resolution on the order of an arcsecond, seeing allowing. The selected science apertures are 1, 2, 5, 10, 16, 30 and 55 arcseconds. For reference, 16 arcseconds corresponds to the angular size of Earth as seen from the distance of the Sun. Owing to their small physical dimensions (from 35 µm to 2 mm in diameter) and in accordance with our radiometric model, the selected telescope aperture is required to ensure enough light reaches ESPRESSO in all configurations.

Two simple refractors, the 'science' and the 'imaging' telescopes, known jointly as the pointing telescope (PT), are piggybacked onto the MT. Light from both the MT and the PT will be collected by the respective frontends and, via fibre links, delivered to the ESPRESSO spectrograph via its calibration unit.

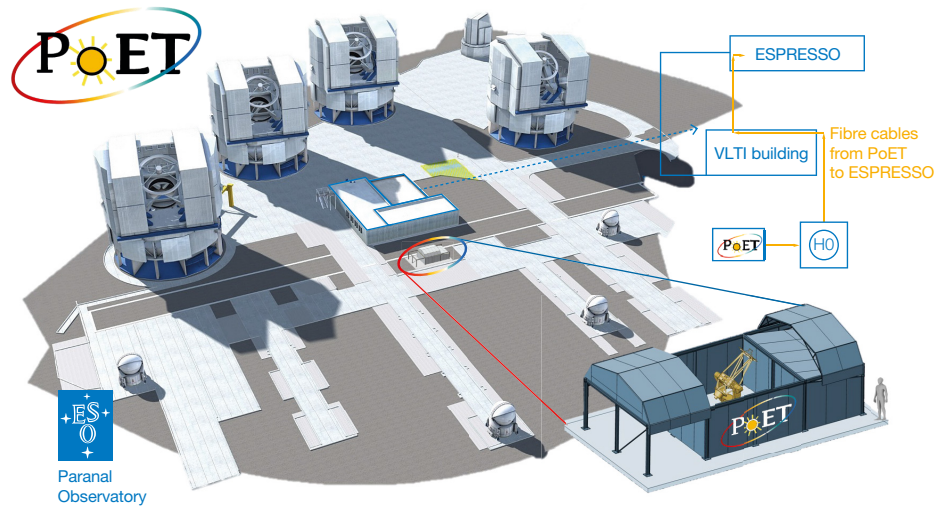

The PoET frontends

As depicted in Figure 4 (right), in the frontend of the MT, the f/12 beam passes through a 90:10 beam-splitter and is directed to the aperture selector by a folding mirror. The set of different apertures are positioned in the focal plane where light is fed into small fibre patches, either directly (apertures from 1 to 5 arcseconds), through demagnifying lenses (from the remaining apertures) or, alternatively, using an integrating sphere (for the 55-arcsecond mode). The selection is made using a translation stage. The light from the exit of the selected aperture is injected into a double scrambler to guarantee a homogeneous distribution into the science fibre that connects with ESPRESSO.

The apertures are metallic pinholes that reflect part of the incident light back to the beam-splitter and redirect it to a guiding camera (imaging system and sensor), where an image of the observed Sun (with a 3-arcmin-diameter field of view) is registered. This image will indicate the location of the observed region and can be seen with either a 1-nm FWHM H$\alpha$ filter or a 10-nm red filter (centred on the same region).

The pointing telescope is comparatively much simpler in design (Figure 4, left). A science refractor telescope (achromat lens), with an aperture of 60 mm and focal length of 100 mm, will image the Sun into a 10-mm-diameter integrating sphere that injects the light into the 'Sun's disc-integrated science fibre', connecting to

Figure 2. Left: Illustration of the location of PoET in Paranal. Upper-right: Schematic of science fibres path from PoET to ESPRESSO (one floor below). Lower-Right: Dome concept with main telescope.

ESPRESSO. A second refractor, with a 50-mm-diameter achromat with a focal length of 350 mm, images the full Sun in the pointing camera sensor. The guiding and pointing cameras will be calibrated to allow the identification of the area to be observed by the MT.

Link to ESPRESSO

The instrument will be installed behind the Very Large Telescope Interferometer (VLTI) building and after the 'no-build zone' (Figure 2). This location was chosen as the preferred site for the instrument based on its relative proximity to ESPRESSO and relatively low wind speeds: to the north, the predominant wind direction, it is 'protected' by the VLTI building, which avoids issues related to vibrations of the telescope structure.

The concept for the telescope dome is a 'sliding roof' design. The roof consists of two sections that open in approximately the west–east direction. The telescope will be hosted in a 5 m × 5 m space. An additional room will be used to house the electronics cabinets, computers, and other auxiliary items.

Science fibres, communication, and power cables will be routed into an unused VLT Auxiliary Telescope hatch (H0), allowing easier access for the fibres towards ESPRESSO through the VLTI tunnel, with minimal structural impact. Based on the observatory architectural drawings, we project around 80 m of cable for each science fibre, needed to deliver light collected by PoET towards ESPRESSO's calibration unit, where PoET fibres will be connected. Inside the dome, the science fibres are split via a set of collimators that allow the insertion of a blue filter. Depending on the scientific requirement, this filter can be selected to compensate for the signal loss in the blue region of the spectra, mainly caused by the fibre attenuation.

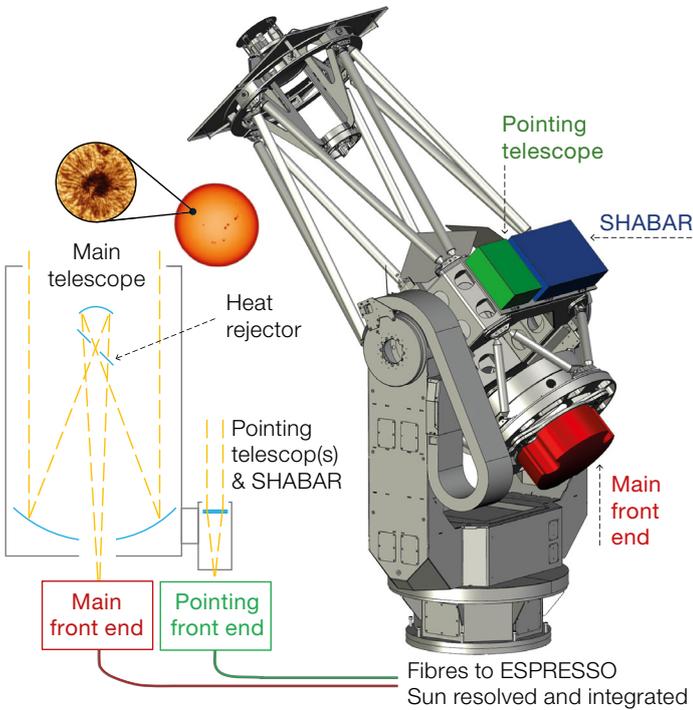

Figure 3. Concept of PoET (left) and scheme of the main telescope (right) with the positioning of the two frontends and the SHABAR seeing measurement device.

### SHABAR

The MT will receive support from a custom instrument (known as SHABAR) to measure the daylight seeing conditions. Seeing is the effect of random fluctuations in the index of refraction throughout Earth's atmosphere, resulting in random fluctuations in the direction of light from a distant source. Scintillation is the random fluctuation of the light intensity received. A correlation was found between the two. With this concept, and making use of a non-telescopic method, we can measure seeing during the day, using the Sun as the source of light. For this purpose, we built a seeing measurement device, following the same concept as Sliepen et al. (2010), to be used along with PoET. Our SHABAR (Wehbe et al., 2024) is currently commissioned and being tested to be ready for PoET's first light.

### Observations, operations and data

As a visitor telescope at ESO's La Silla–Paranal Observatory, PoET is expected to be fully autonomous and not require any local intervention. As such, PoET's software will oversee all operations of the telescope and allow it to run either in a fully autonomous mode or managed remotely from our premises in Portugal. Typically, after the daily calibrations, science operations will run a short script on the ESPRESSO workstation to prepare the instrument to execute PoET–ESPRESSO observations. On its side, the PoET software will create science Observation Blocks (OBs) and ingest them into the ESPRESSO execution sequence through the P2 Application Programming Interface[2]. From then on, and until science operations require ESPRESSO for the preparation of the night operations, PoET–ESPRESSO observations will be executed automatically.

Observations will be made in both high-resolution and ultra-high-resolution modes, at spectral resolutions of ~140 000 and ~200 000, respectively (Pepe et al., 2021). The baseline exposure time used to define the instrumental design is 30 seconds, a compromise between the interest in resolving the solar oscillation signals and the overhead time related to the readout of the ESPRESSO detector.

Several observing modes were defined, leveraging the optical setup of the PoET–ESPRESSO interface and PoET's scientific requirements. The operation modes can be divided into the following categories:

1. Observations of the resolved Sun: in this mode, PoET will acquire high-resolution spectra of a selected (resolved) region of the solar disc using the MT;
2. Resolved solar disc observations with simultaneous 'Sun-as-a-star': in this mode, both the PT and the MT will be injecting light into ESPRESSO;
3. Sun-as-a-star observations with simultaneous wavelength calibration: meant to acquire high-precision radial velocities time-series from disc-integrated exposures ('Sun-as-a-star'), mimicking the usual stellar observations of ESPRESSO.

In all cases, disc-integrated observations will feed ESPRESSO's fibre A, while disc resolved observations will feed fibre B. As a consequence, it will not be possible to obtain simultaneous wavelength calibrations while observing in the disc-resolved mode. This will not impact the science goals of PoET, as ESPRESSO drifts are a few orders of magnitude smaller than the local velocities of the solar disc (for example, due to granulation).

PoET data will be reduced with a new version of the standard ESPRESSO Data Reduction Software (DRS). The DRS will produce the usual ESPRESSO science-grade data products for the different observing modes detailed above. The reduced data will then be complemented with an auxiliary FITS file containing context images from the PoET cameras as well as other relevant information (for example, the seeing measured with SHABAR). All data products will be uploaded to the ESO Phase 3 archive.

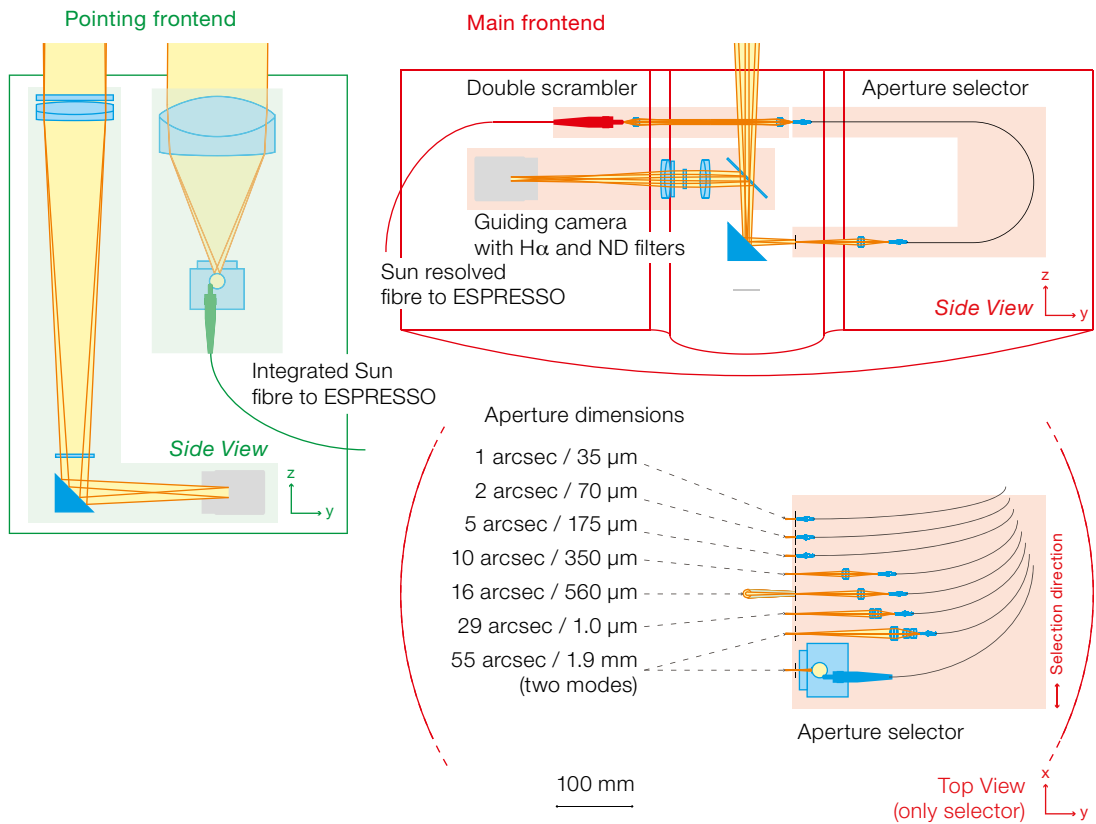

Figure 4. Concept of the two frontends and the fibre injection system and its aperture selector.

### Timeline and expected science impact

The detailed design of the PoET telescope is now concluded. The procurement of the main components, telescope and mount, as well as the dome, is also in progress. We expect to install the telescope on Paranal in the second semester of 2025 and start operations soon after.

The main scientific motivation behind PoET is to tackle the problem of 'stellar noise' in high-resolution spectroscopic observations, both in precise radial velocities and in transmission/emission spectroscopy. This will be fundamental to the success of present and future efforts in exoplanet research, including those linked with ELT instrumentation (for example, ANDES) and ESA missions (for example, PLATO). Moreover, PoET is presently raising interest among several scientific communities, and its data are expected to contribute to other science cases, including solar and stellar physics.

### Acknowledgements

We would like to acknowledge the fruitful discussions with the scientific community at the PoET Workshops[3] organised in 2023 and 2024, that helped to define the final design of the instrument and the planning of the scientific observations. We would like to express our gratitude to the staff at Officina Stellare for their support in the development of the solar telescope. The project is funded by the European Union (ERC, FIERCE, 101052347). Views and opinions expressed are, however, those of the author(s) only and do not necessarily reflect those of the European Union or the European Research Council. Neither the European Union nor the granting authority can be held responsible for them. This work was supported by the Fundação para a Ciência e a Tecnologia (FCT) through national funds by these grants: UIDB/04434/2020 DOI: 10.54499/UIDB/04434/2020, UIDP/04434/2020 DOI: 10.54499/UIDP/04434/2020, and the PhD grant UI/BD/152077/2021 DOI: 10.54499/UI/BD/152077/2021.

### Links

[1] PoET web page: https://poet.iastro.pt
[2] ESO Phase 2 API: https://eso.org/sci/observing/phase2/p2intro/Phase2API.html
[3] PoET workshops: https://poet.iastro.pt/events/